\documentclass[10pt,letterpaper,twocolumn,english,conference]{IEEEtran}
\usepackage[T1]{fontenc}
\usepackage[latin9]{inputenc}
\usepackage{color}
\usepackage{units}
\usepackage{amsmath}
\usepackage{graphicx}
\usepackage[numbers]{natbib}

\makeatletter

\pdfpageheight\paperheight
\pdfpagewidth\paperwidth

\makeatother

\usepackage{babel}
\begin{document}

\title{\textcolor{black}{\huge MAC Centered Cooperation - Synergistic Design
of Network Coding, Multi-Packet Reception, and Improved Fairness to
Increase Network Throughput}\textcolor{black}{\thanks{This work is sponsored by the Department of Defense under Air Force  Contract FA8721-05-C-0002. Opinions, interpretations, recommendations, and conclusions are those of the authors and are not necessarily endorsed by the United States Government. Specifically, this work was supported by Information Systems of ASD(R\&E). Contributions of the Irwin Mark Jacobs and Joan Klein Jacobs Presidential Fellowship have also been critical to the success of this project.}}}

\author{\textcolor{black}{\small Jason Cloud{*}$^{\dagger}$, Linda Zeger$^{\dagger}$,
Muriel Médard{*}}\\
\textcolor{black}{\small {*}Research Laboratory of Electronics, Massachusetts
Institute of Technology, Cambridge, MA.}\\
\textcolor{black}{\small $^{\dagger}$MIT Lincoln Laboratory, Lexington,
MA.}\\
\textcolor{black}{\small Email: \{jcloud@, zeger@ll., medard@\}mit.edu}}
\maketitle
\begin{abstract}
\textcolor{black}{We design a cross-layer approach to aid in developing
a cooperative solution using multi-packet reception (MPR), }network
coding (NC)\textcolor{black}{, and medium access (MAC). We construct
a model for the behavior of the IEEE 802.11 MAC protocol and apply
it to key small canonical topology components and their larger counterparts.
The results obtained from this model match the available experimental
results with fidelity. Using this model, we show that fairness allocation
by the IEEE 802.11 MAC can significantly impede performance; hence,
we devise a new MAC that not only substantially improves throughput,
but provides fairness to }\textit{\textcolor{black}{flows}}\textcolor{black}{{}
of information rather than to }\textit{\textcolor{black}{nodes}}\textcolor{black}{.
We show that cooperation between }NC\textcolor{black}{, MPR, and our
new MAC achieves super-additive gains of up to 6.3 times that of routing
with the standard IEEE 802.11 MAC. Furthermore, we extend the model
to analyze our MAC's asymptotic and throughput behaviors as the number
of nodes increases or the MPR capability is limited to only a single
node. Finally, we show that although network performance is reduced
under substantial asymmetry or limited implementation of MPR to a
central node, there are some important practical cases, even under
these conditions, where MPR, }NC\textcolor{black}{, and their combination
provide significant gains.\IEEEpeerreviewmaketitle}

\textcolor{black}{}
\end{abstract}

\section{\textcolor{black}{\thispagestyle{empty}Introduction}}

\textcolor{black}{With the increase in wireless use, current wireless
systems are throughput limited and are difficult to scale to large,
dense networks. We develop a simple model that is easily extended
to analyze the asymptotic regime and asymmetric traffic so that we
can evaluate the performance of combining various techniques to increase
network throughput and reduce overall delay.}

\begin{figure}
\begin{centering}
\textcolor{black}{\includegraphics[width=3in]{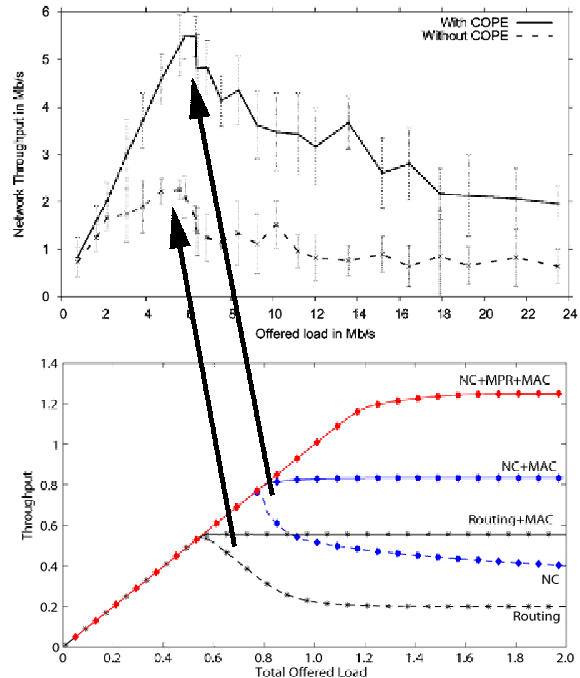}}
\par\end{centering}

\textcolor{black}{\caption{Comparison of the empirical COPE performance data collected from a
20-node IEEE 802.11 wireless ad hoc network test bed (top), \citep{Katti01},
and the resulting throughput using a model of the IEEE 802.11 MAC
proposed by \citep{ZhaoF01} (bottom). This model is th\textcolor{black}{e
starting point for o}ur analysis with MPR and development of our improved
MAC. \label{fig:Emperical-COPE-performance}}
}
\end{figure}

Network coding (NC), introduced\textcolor{black}{{} by \citep{Ahlswede01},
and the proof that simple, linear network codes can achieve the multicast
capacity by \citep{Koetter01} led to a new heuristic inter-session
NC scheme, Coding Opportunistically (COPE), for wireless networks.
Proposed by Katti }\textsl{\textcolor{black}{et. al.}}\textcolor{black}{{}
\citep{Katti01}, COPE is a cooperative }NC\textcolor{black}{{} scheme
that identifies coding opportunities and exploits them by forwarding
multiple packets in a single transmission. While \citep{Koetter01}
showed that inter-session network coding is generally very difficult,
COPE circumvents these complexity issues by decoding at each hop.
The use of this simple coding scheme was shown to provide up to 3
to 4 times the throughput capacity over simply routing packets through
the network. Implementing COPE in a 20-node IEEE 802.11 test bed,
Katti }\textit{\textcolor{black}{et. al}}\textcolor{black}{. provided
empirical data, shown in the upper half of Fig. \ref{fig:Emperical-COPE-performance},
that shows the benefits of using COPE in wireless mesh networks.}

Sengupta \textit{et. al.}, \citep{Sengupta01} and Le \textit{et.
al.}, \citep{Le01} provided analyses of these experimental results,
but only considered coding a maximum of two packets together at a
time and did not address the interaction between NC and the medium
access (MAC) fairness. \textcolor{black}{As a result, their analyses
provide throughput gains that are considerably smaller than the experimental
results and do not explain the non-monotonic behavior of the results
shown in Fig. \ref{fig:Emperical-COPE-performance}. Zhao and Médard,
\citep{ZhaoF01}, modeled the same experimental results, but showed
that the }\textit{\textcolor{black}{fairness}}\textcolor{black}{{} imposed
by the IEEE 802.11 MAC explains this non-monotonic behavior. In addition,
they demonstrated that the majority of the throughput gain achieved
by using COPE is a result of coding three or more uncoded, or native,
packets together at time. They showed that these gains are not reflected
in three node network models, used in prior analyses, and at least
five nodes are required to accurately capture the throughput gains
from }NC\textcolor{black}{. The NC and routing curves in Fig. \ref{fig:Emperical-COPE-performance}
show that the results obtained using their model for a simple 5-node
cross }component\textcolor{black}{{} \citep{ZhaoF01} is consistent
with the empirical data from \citep{Katti01}. Furthermore, Seferoglu
}\textit{\textcolor{black}{et. al.}}\textcolor{black}{{} \citep{Seferoglu01}
used this 5-node }component\textcolor{black}{, and variants of them,
to analyze TCP performance over coded wireless networks. Hence, we
consider the 5-node cross }component\textcolor{black}{{} and additional
5-node }component\textcolor{black}{s, as well as their extensions
to any number of nodes, in order to understand the effects of combining
}NC\textcolor{black}{{} and multi-packet reception (MPR) in larger networks.}

\textcolor{black}{While the performance of COPE significantly increases
network throughput \citep{Katti01}, it does not completely alleviate
multi-user interference. With the development of new radio technologies,
the ability to receive multiple packets simultaneously at the physical
layer makes it possible to increase throughput and also has the potential
to reduce contention among users \citep{Garcia-Luna-Aceves02}. The
stability of slotted ALOHA with MPR, but not NC, was studied by \citep{Ghez01},
and several protocols implementing MPR have been proposed by \citep{ZhaoQ01}
and \citep{Celik01}. However, little analysis has been performed
in evaluating schemes that use both MPR and }NC\textcolor{black}{.
Garcia-Luna-Aceves}\textit{\textcolor{black}{{} et al.}}\textcolor{black}{{}
\citep{Garcia-Luna-Aceves01} }\textit{\textcolor{black}{compared}}\textcolor{black}{{}
the use of }NC\textcolor{black}{{} to MPR, but did not consider their
}\textit{\textcolor{black}{combined}}\textcolor{black}{{} use. In addition,
Rezaee }\textit{\textcolor{black}{et al.}}\textcolor{black}{{} \citep{Rezaee01},
provided an analysis of the combined use of }NC\textcolor{black}{{}
and MPR in a fully connected network, but did not consider the effects
of bottlenecks or multi-hop traffic.}

\textcolor{black}{We provide an analysis of the combined use of }NC\textcolor{black}{{}
and MPR in a multi-hop, congested network. We extend the initial model
proposed by \citep{ZhaoF01} to include various topology configurations,
asymmetric and asymptotic behavior, and various implementations of
MPR in order to show that the achievable throughput when using }NC\textcolor{black}{{}
in conjunction with MPR in a cooperative}\textit{\textcolor{black}{{}
}}\textit{\textcolor{black}{\emph{multi-hop}}}\textcolor{black}{{} network
is }\textcolor{black}{\emph{super-additive}}\textcolor{black}{. We
then use this model to design a cross-layer solution that increases
throughput subject to the constraint of fairness between flows, rather
than between nodes, for congested network structures. While MAC fairness
has been previously studied \citep{Abuzanat01}, our solution uses
cooperation between nodes and takes into account the interaction among
MPR, }NC\textcolor{black}{, and MAC. Using our simplified model, we
then analyze the behavior of our solution as the traffic across the
bottleneck becomes asymmetric, as well as in the asymptotic regime
as the number of nodes in each }component\textcolor{black}{{} increases.
Finally, we analyze the throughput behavior as we limit the MPR capability
to a subset of nodes within the network.}

\textcolor{black}{The remainder of the paper is organized as follows:
Section \ref{sec:Network-models} describes the network models used
in our analysis. Section \ref{sec:Multi-Packet-Reception-and} provides
an analysis of }NC\textcolor{black}{{} and MPR for 5-node network }component\textcolor{black}{s
with the current IEEE 802.11 MAC fairness allocation. Section \ref{sec:Improving-the-MAC}
demonstrates the importance of considering the MAC when using a combined
MPR and }NC\textcolor{black}{{} solution and provides an improved MAC
that increases throughput while ensuring fairness to flows of information
rather than to nodes. Sections \ref{sec:Asymmetric-Traffic} and \ref{sec:Performance-of-large-N}
investigate the effects of asymmetric network traffic and the gains
obtained when considering delay in the asymptotic regime, respectively,
with the new MAC. Section \ref{sec:Limited-MPR} provides an analysis
of the throughput when the MPR capability is limited to a subset of
nodes. Finally, we conclude with a comparison of the results in Section
\ref{sec:Conclusion}.}

\section{\textcolor{black}{Network Models and Parameters\label{sec:Network-models}}}

This section develops a simple model that gives insight into cross-layer
design of wireless networks by using NC, various MAC approaches, and
MPR. We identify each network element's fundamental behavior and model
them using simple, intuitive methods so that various performance measures
can be evaluated and design trade-offs can be weighed. Subsequent
sub-sections identify specific behaviors of these elements and describe
the abstractions and simplifications needed to make the model tractable.

The scenario considered consists of a wireless error-free packet network
that is operated in fixed-length time-slots. Each node is half-duplex
(i.e., cannot receive and transmit in the same time-slot), and only
one packet can be sent per time-slot by any given node. If multiple
packets are received by a node in the same time-slot without using
MPR, it is assumed that a collision occurs and all packets are lost.

\subsection{Network Topologies}

\begin{figure}
\begin{centering}
\textcolor{black}{\includegraphics[width=2.75in]{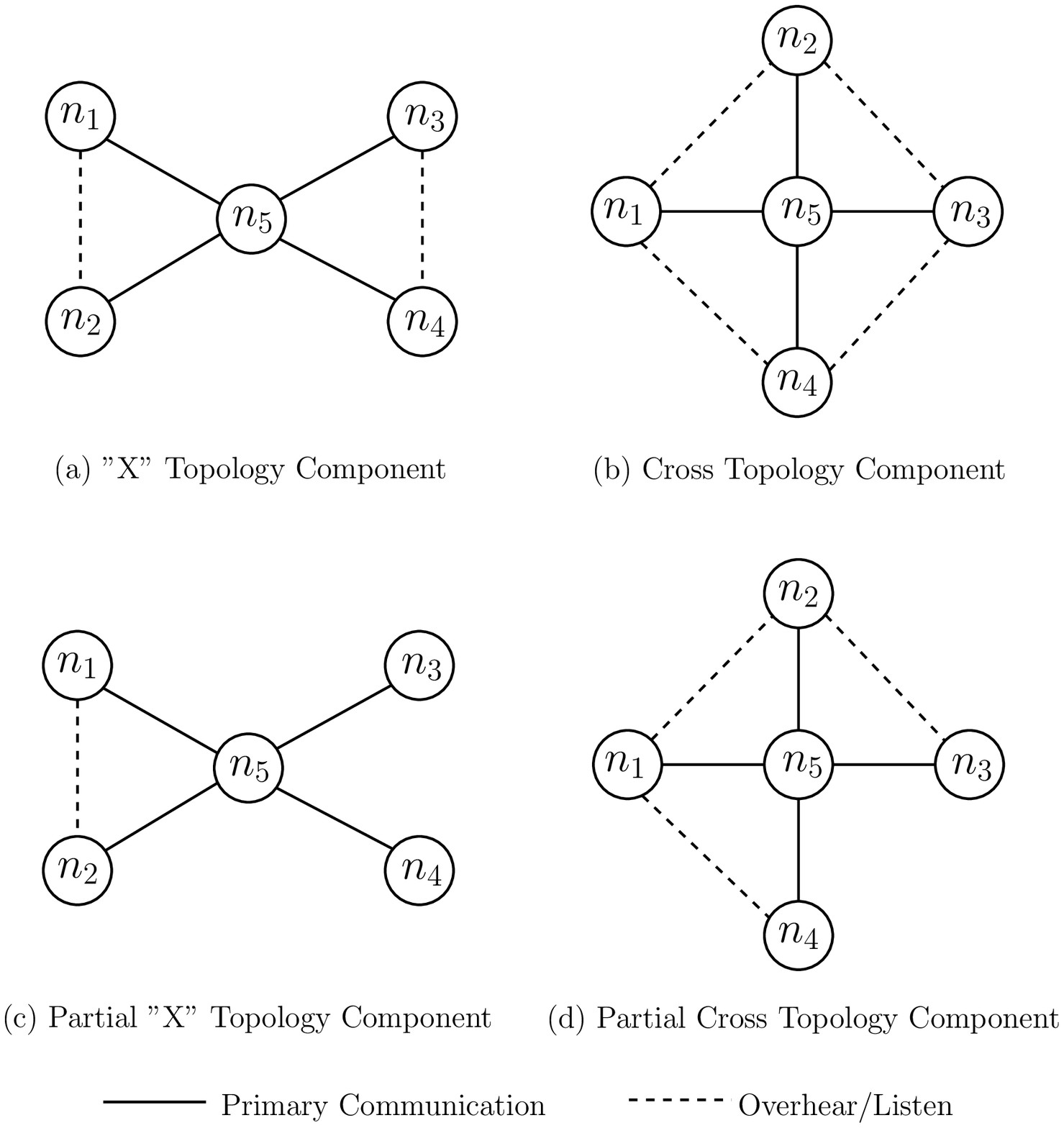}}
\par\end{centering}

\textcolor{black}{\caption{\label{fig:Network-Topologies}\textcolor{black}{Basic network structures
responsible for traffic bottlenecks and congestion in larger networks.
We analyze these components and variants of them.}}
}
\end{figure}

\textcolor{black}{Our model uses the five node canonical }component\textcolor{black}{s
shown in Fig. \ref{fig:Network-Topologies}. These components are
of interest for two reasons. First, they form the primary structures
in larger multi-hop networks that create bottlenecks and congestion.
By looking at the traffic that travels through the center node, these
components help us model the performance gains of multi-hop traffic
under both low and high loads. Second, the COPE experimental results
show that the majority of the coded packets generated in the network
contained, on average, 3-4 native packets \citep{Katti01}. As a result,
each component used must then be of sufficient size to capture the
majority of the gains seen in \citep{Katti01}. The canonical topology
components in Fig. \ref{fig:Network-Topologies} reflect all of the
possible combinations of five node multi-hop networks that allow for
the potential coding of up to four unencoded, or native, packets.}

\textcolor{black}{Each component has specific constraints due to its
structure and will affect the performance of the MAC, }NC\textcolor{black}{,
and MPR in different ways. In Fig. \ref{fig:Network-Topologies},
we define these constraints through the use of a solid edge that depicts
active, or primary communication, and a dotted edge that depicts passive,
or overhear/listening communication. The absence of an edge between
any two nodes indicates that all communication between the two nodes
must be routed through the center node. The center node $n_{5}$ in
each component is fully connected regardless of the topology, and
traffic flows originating from the center require only a single hop
to reach their destination. Within the ``X'' and partial {}``X''
}component\textcolor{black}{s, all flows originating from a node in
a given set terminate at a node in the opposite set; and within the
cross and partial cross }component\textcolor{black}{,s each traffic
flow originating from a given node is terminated at the node directly
opposite the center. For example, nodes $n_{1}$, $n_{2}$, and $n_{5}$
in the {}``X'' }component\textcolor{black}{{} are fully connected
and nodes $n_{3}$, $n_{4}$, and $n_{5}$ are also fully connected;
but $n_{1}$ and $n_{2}$ are not connected to $n_{3}$ and $n_{4}$.
All traffic between any node $\{n_{1},n_{2}\}\in X_{1}$ and any node
$\{n_{3},n_{4}\}\in X_{2}$ must travel through the center.}
\textcolor{black}{}
\begin{figure}
\begin{centering}
\includegraphics[width=2.5in]{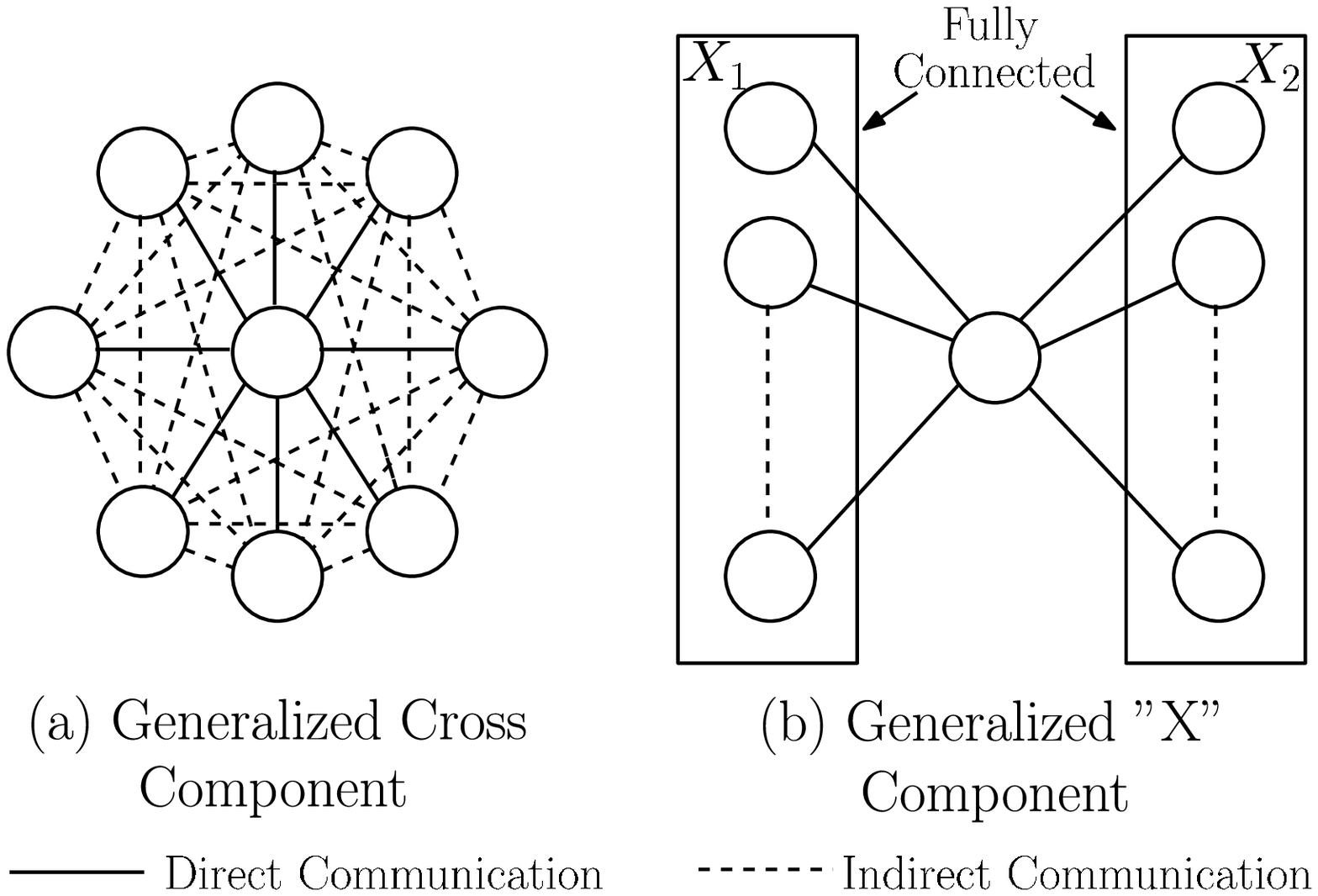}
\par\end{centering}

\textcolor{black}{\caption{\textcolor{black}{\label{fig:Generalized-topology-components}Generalized
topology components for $N$ nodes.}}
}
\end{figure}

The study of topology components extended to an arbitrary number $N$
of transmitting nodes, $n_{i}\in{\cal N}$ where $i\in[1,N]$, aids
in the analysis of performance and delay in larger networks. Sections
\ref{sec:Improving-the-MAC} and \ref{sec:Performance-of-large-N}
use the variants of the cross and {}``X'' components shown in Fig.
\ref{fig:Generalized-topology-components} to provide insight into
the achievable gains and cross-layer design of networks employing
the various technologies described here. For the cross component,
there are $N-1$ transmitting edge nodes and a single center, or relay,
node. All edge nodes are connected with the center node and connected
with all other edge nodes except the one directly opposite the center.
Each node generates traffic destined only for the node directly opposite
the center. For the \textquotedblleft{}X\textquotedblright{} component,
there are also $N-1$ transmitting edge nodes and a single center
node. The edge nodes are split into two sets, $X_{1}$ and $X_{2}$.
All edge nodes within a given set are fully connected and are also
connected to the center. Each node generates traffic destined for
a node within a different set. Furthermore, it is assumed that the
set of transmitting nodes $\mathcal{N}$ is stable and does not frequently
change. This eliminates the need to consider a decision mechanism
for determining which nodes are transmitting.

\subsection{Network Coding Model}

\textcolor{black}{Network coding is modeled by considering the ability
of a given node to combine multiple packets together. We use} COPE
\citep{Katti01} as a case study. COPE inserts a coding shim between
the IP and MAC layers and uses the broadcast nature of the wireless
channel to opportunistically code packets from different nodes using
a simple XOR operation.\textcolor{black}{{} The wireless channel enables
each node to overhear packets that can be used to help decode any
subsequently received encoded packets.}

\textcolor{black}{In the proposed model,} each encoded packet is sent
if and only if it can be decoded by the intended recipients (i.e.,
the intended recipients have overheard enough native packets, or degrees
of freedom, to enable each encoded packet to be decoded). In addition,
only the center node will encode packets together while each edge
node will always transmit their packets unencoded due to the limitations
imposed by the \textcolor{black}{components in Figures \ref{fig:Network-Topologies}
and \ref{fig:Generalized-topology-components}. }The model further
assumes that feedback is perfect and that each node knows the native
packets overheard by its neighbors. Consistent with COPE's implementation,
each packet is sent as a broadcast transmission on the channel at
the first opportunity without delay and each information flow does
not exercise congestion control (i.e., each packet generated is part
of a UDP session). Finally, \textcolor{black}{neither the complexity
of the coding or decoding operations nor any other aspects of the
}NC\textcolor{black}{{} implementation found in \citep{Katti01} are
considered since their contributions to the overall network performance
is small.}

\subsection{IEEE 802.11 MAC Model}

\textcolor{black}{We model the IEEE 802.11 MAC's distributed coordination
function (DCF) \citep{80211MAC}, which uses carrier sense multiple
access with collision avoidance (CSMA/CA) as the method in which a
node accesses the channel. In order to develop the model, we first
identify the behavior of the DCF over a sufficiently long period of
time. The non-monotonic behavior observed in the COPE experiments
shown in Fig. \ref{fig:Emperical-COPE-performance} was noted by \citep{ZhaoF01}
to be a result of the IEEE 802.11 MAC fairness mechanism, which essentially
distributes channel resources equally among competing nodes. This
realization is also consistent with the analysis and simulation results
presented in \citep{Duda01}, which showed that the probability of
a node successfully accessing the channel converges to $\nicefrac{1}{N}$
for $N$ competing nodes. The MAC model used for each non-MPR case
below captures this limitation with fidelity by limiting each node's
access to the channel to $\nicefrac{1}{N}$ of the time when the channel
is fully saturated, while the random access protocols are simplified
so they match the experimental throughput behaviors found in \citep{Katti01}.
For example, the non-monotonic behavior in Figure \ref{fig:Emperical-COPE-performance}
is a result of both collisions and fairness; but the total effects
of collisions on throughput from either hidden nodes or identical
back-off times are small in relation to the effects of the IEEE 802.11
MAC fairness mechanisms. When using MPR, extensions to the MAC model
are required where Section \ref{sub:Multi-Packet-Reception-Model}
explains these extensions and the reasons for their necessity.}

\textcolor{black}{The MAC model also assumes that the request-to-send/clear-to-send
(RTS/CTS) feature of the DCF is not used, which is consistent with
the IEEE 802.11 ad-hoc mode and the COPE experiments \citep{Katti01}.
Furthermore, the model does not consider the additional effects on
overall throughput associated with various implementation aspects
of the DCF. Since the DCF introduces a constant overhead that lowers
the throughput to about 20\% to 30\% of the bit rate depending on
the variant of 802.11 used \citep{Duda01}, these assumptions provide
upper bounds to the achievable throughput. In addition, implementing
MPR and NC reduces the number of times each node is required to access
the channel and therefore reduces the overhead incurred by the DCF.
This results in tighter throughput bounds than when MPR and NC are
not used and also provides an estimate for the MPR and NC throughput
gains that can be achieved. Finally, any additional time needed to
acknowledge transmitted packets is included in the duration of each
of the model's time-slots. This allows for a new packet to be transmitted
in each integer time-slot.}

\subsection{Multi-Packet Reception Model\label{sub:Multi-Packet-Reception-Model}}

In general, MPR allows for the correct reception at the physical layer
of one or more packets involved in a collision. Several techniques
can be used to implement MPR in a wireless network, for example: Code
Division Multiple Access (CDMA), Space Division Multiple Access (SDMA),
Orthogonal Frequency Division Multiple Access (OFMDA), etc. The fundamental
concept behind each of these technologies is that a receiver is able
to separate signals transmitted simultaneously from different nodes
and then extract the required data from each transmission.

The analysis and simulations in the remainder of the paper will evaluate
the potential throughput gains using two MPR models. In both, the
number of simultaneous transmissions that a node can successfully
receive without a collision is $m$ where only $m=\{1,2,4\}$ are
considered in this paper. In the first model, CSMA/CA is strictly
enforced for $m=\{1,2\}$. If a node senses any other node transmitting,
it will follow the 802.11 DCF algorithm and not attempt to transmit
until the channel is idle again. This model essentially uses MPR to
minimize the hidden terminal problem. When $m=4$, a slight generalization
of the traditional CSMA/CA is required. \textcolor{black}{We pick
the combination of transmitting nodes such that the average number
of transmissions received by any given node within the network is
maximized.} It is important to note that this generalization allows
strictly fewer than four \emph{adjacent} nodes to transmit at the
same time. For example, consider the generalized {}``X'' component
shown in Fig. \ref{fig:Generalized-topology-components}. We will
only allow two nodes in set $X_{1}$ and only two nodes in set $X_{2}$
to transmit at the same time, which maximizes the number of packets
that each edge node overhears. In the second model (referred to as
MPR-adapted CSMA in the remainder of the paper), a node will be allowed
to transmit as long as the number of simultaneous transmissions sensed
is less than $m$. For the generalized {}``X'' component, this model
will allow for up to four nodes within a single set to transmit at
the same time. In either model, up to $m$ packets can be sent simultaneously
in the same time-slot. 

When considering the IEEE 802.11 MAC model developed in the previous
sub-section, the fraction of time each node accesses the channel under
saturated conditions is dependent on the MPR model and component used.
In general, the time each node will be able to access the channel
as it saturates becomes approximately $\nicefrac{1}{\left(\left\lceil \frac{N-1}{m}\right\rceil +1\right)}$
where the MPR model and component used will dictate the exact allocation
of channel resources among the set of nodes. Specific details and
additional explanation will be expanded upon in later sections.

\subsection{Additional Model Assumptions and Parameters}

The channel is divided into 100 time-slots where each time-slot uses
$\nicefrac{1}{100}$ of the total amount of channel resources available
to the $N$ transmitting nodes. Successful transmission of each packet
requires a full time-slot therefore requiring $\nicefrac{1}{100}$
of the total amount of channel resources. Performance is evaluated
at various values of $k_{T}\in\left[1,200\right]$ where $k_{T}$
is defined as the total number of packets in the network and is deterministic.
In order to model stochastic packet arrivals, these $k_{T}$ packets
are distributed to each node where each node has $K_{i}$, $i\in[1,N]$,
packets and $\left(K_{1},K_{2},\ldots,K_{N}\right)$ is distributed
according to a joint binomial distribution, given $k_{T}$ and $N$,
with parameters $n_{i}=k_{T}-\sum_{j=1}^{i-1}k_{i}$ and $p_{i}=(N-i+1)^{-1}$.
The number of packets each node has to send will be referenced in
later sections as the fraction of the total channel resources, or
load $\rho_{i}$, required to send all $k_{i}$ packets one hop (i.e.,
$\rho_{i}=\nicefrac{k_{i}}{100}$). In addition, the total offered
load $P$ to the network is deterministic, given $k_{T}$, and is
defined as $P=\sum_{i\in N}\rho_{i}=\nicefrac{k_{T}}{100}$.

In order to determine specific network behaviors over the range of
$k_{T}$, we define the total network component load $P_{T}$ as the
load induced in the network component as a result of NC, MPR, and
MAC. This allows us to specify three regimes which are of particular
interest. These regimes are: the unsaturated throughput regime ($P_{T}<1$),
the maximum throughput regime ($P_{T}=1$), and the saturated throughput
regime ($P_{T}>1$). In general, the component load $P_{T}=P$ when
routing packets without NC and/or MPR and $P_{T}\leq P$ otherwise.
Specifically, the component load $P_{T}$ is a random variable, with
sample value $p_{T}$, that is defined as the sum of the load $L_{R}$
induced by relaying packets through the center node $n_{center}$,
and the load $L_{M}$ required to send each native packet one-hop
(i.e., $P_{T}=L_{R}+L_{M})$. The sample values, $l_{R}$ and $l_{M}$,
for $L_{R}$ and $L_{M}$ respectively are bounded by:

\begin{equation}
\frac{1}{c}\sum_{i\in\mathcal{N}\setminus n_{center}}\rho_{i}\leq l_{R}\leq\sum_{i\in\mathcal{N}\setminus n_{center}}\rho_{i},\label{eq:Relay-Load}
\end{equation}

\begin{equation}
\frac{1}{m}\sum_{j\in\mathcal{N}\setminus n_{center}}\rho_{j}+\rho_{center}\leq l_{M}\leq\sum_{j\in\mathcal{N}}\rho_{j},\label{eq:One-hop-load}
\end{equation}
where the coefficient $c$ is the number of packets that can be encoded
together by $n_{center}$, and $\rho_{center}$ is the fraction of
time, or load, needed to send all of the packets originating at $n_{center}$
one-hop. The relay load $L_{R}$ is a function of the number of packets
that can be encoded together by $n_{center}$ and only counts the
load required to send relayed packets a second hop. The one-hop load
$L_{M}$ consists of the load needed to send all of the edge node's
packets to $n_{center}$, which is a function of $m$, and the load
$\rho_{center}$ required by $n_{center}$ to send its own packets
to the edge nodes. The lower bounds for each are functions of the
component's configuration as well as the difference in each node's
initial load. Each lower bound is met with equality if each $\rho_{i}=\rho_{j}$,
$i,j\in{\cal N}\setminus n_{center}$, and $i\neq j$. The upper bound
in eq. (\ref{eq:Relay-Load}) is met with equality if no coding opportunities
occur at $n_{center}$ and in eq. (\ref{eq:One-hop-load}) if no simultaneous
transmissions occur. Given sample values of each node's load $\rho_{i}$,
$i\in N$, and the component, both $L_{R}$ and $L_{M}$ are deterministic.
Section \ref{sec:Multi-Packet-Reception-and} will provide additional
clarification and examples.

Furthermore, the allocated load $S_{i}$ is defined as the amount
of channel resources given to each node in the network as a result
of the MAC. \textcolor{black}{When $P_{T}\leq1$, each node is allocated
enough time-slots to send all of its packets through the component.
The allocated load in this case is $s_{i}=\rho_{i}$ for $i\in\mathcal{N}\setminus n_{center}$
and the load allocated to the center node is the sum of the load originating
from the center $\rho_{center}$ and the load resulting from relaying
packets (i.e., $s_{center}=\rho_{center}+l_{R}$). As the MAC saturates
(i.e., $P_{T}>1$) the allocated load for each node is $s_{i}\leq l_{i}$,
$i\in{\cal N}\setminus n_{center}$, and $s_{center}\leq\rho_{center}+l_{R}$.}

Finally, t\textcolor{black}{he throughput $S$ is defined in relation
to the number of packets that reach their respective sinks within
the }component\textcolor{black}{. For $P_{T}\leq1$, the channel is
not saturated so the MAC does not limit each node from sending all
of its packets. As a result, the throughput $S=P$. For $P_{T}>1$,
the channel is saturated and the MAC must limit the number of transmissions
made by each node in order to remain within the channel constraints
(i.e., the amount of resources allocated to the sum of nodes can be
no more than one or $\sum_{i\in\mathcal{N}}s_{i}\leq1$). The MAC
limits the number of transmissions by adjusting the allocated load
for each node according to the proposed model. In the saturated regime,
the throughput saturates to the amount of information that the center
node can transmit per time-slot. For example, the IEEE 802.11 MAC
will distribute channel resources equally among each transmitting
node and the center node will only receive $\nicefrac{1}{N}$ of the
available resources. The total amount of information that the center
node transmits is then equal to the throughput. Section \ref{sec:Multi-Packet-Reception-and}
will provide greater detail into calculating the throughput with and
without }NC\textcolor{black}{{} and MPR.}

\section{\textcolor{black}{Multi-Packet Reception and Network Coding Performance
Analysis\label{sec:Multi-Packet-Reception-and}}}

\textcolor{black}{With each of the network }component\textcolor{black}{s
shown in Fig. \ref{fig:Network-Topologies}, we analyze the }component\textcolor{black}{{}
performance with and without the use of }NC\textcolor{black}{{} and
MPR. We also consider both unicast and broadcast traffic.}

\subsection{\textcolor{black}{{}``X'' Topology Component Analysis\label{sub:X-Topology-Component}}}

\begin{figure}
\centering{}\textcolor{black}{\includegraphics[width=3.25in]{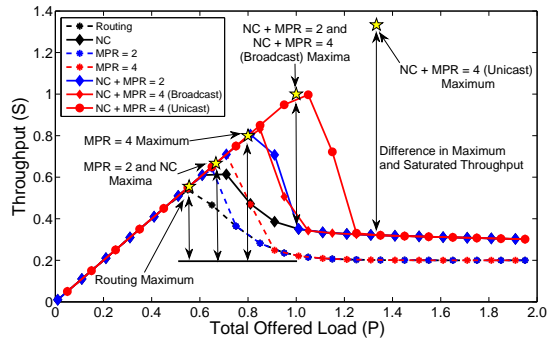}\caption{\label{fig:X-OriginalMAC}Broadcast and unicast throughput for a 5-node
{}``X'' component. The throughput shown is an \emph{average} over
the stochastic packet arrivals. The stars indicate the maximum throughput
which is achieved when there is symmetric full loading of packets.
Each vertical double arrow shows the difference in the maximum and
saturated throughput due to MAC fairness for each case.}
}
\end{figure}

The {}``X'' component, depicted in Fig. \ref{fig:Network-Topologies}(a),
will be used to provide insight into the analysis and to help explain
the simulation results. Fig. \ref{fig:X-OriginalMAC} shows both the
analytical and simulated throughput for each case discussed below.
The stars in the figure indicate the maximum achievable throughput
obtained from analysis when the MPR and/or NC gain is maximized. The
curves show the simulated throughput, which is averaged over the distribution
discussed in Section \ref{sec:Network-models}.\textcolor{black}{}

\subsubsection{\textcolor{black}{Routing (No Network Coding, $m=1$)}}

\textcolor{black}{We will use routing as the baseline for our analysis.
Consistent with the results found in \citep{Katti01} and the analysis
performed in \citep{ZhaoF01}, the throughput increases linearly within
the non-saturated region, $P\in[0,\nicefrac{5}{9})$, and it reaches
its maximum of $S=\nicefrac{5}{9}$, depicted by a star in Fig. \ref{fig:X-OriginalMAC},
when $p_{T}=\sum_{i=1}^{4}\rho_{i}+\sum_{i=1}^{5}\rho_{i}=1$. For
symmetric loads at each node (i.e., $\rho_{i}=\rho_{j}$, $i\neq j$),
each source's individual load is $\rho_{i}=\nicefrac{1}{9}$ for $i\in[1,5]$.}

\textcolor{black}{The throughput saturates for $P>\nicefrac{5}{9}$.
Initially, the IEEE 802.11 MAC allocates time slots to nodes requiring
more resources. The throughput is therefore the amount of time $n_{5}$
is able to transmit, $s_{5}=1-\sum_{i=1}^{4}s_{i}$, which decreases
as $P$ increases. The network component completely saturates when
each node requires a large fraction of the available time-slots and
the MAC restricts each node's access to the channel by ensuring fairness
among all nodes (i.e., $s_{i}=\nicefrac{1}{5}$ for $i\in[1,5]$).
The total saturated throughput is equal to the total amount of information
that $n_{5}$ transmits (i.e., $S=\nicefrac{1}{5}$).}

\subsubsection{\textcolor{black}{Network Coding Only ($m=1$)}}

\textcolor{black}{When using }NC\textcolor{black}{, there are limitations
imposed by the }component\textcolor{black}{'s configuration. Packets
from different nodes within the same set (i.e., $\{n_{1},n_{2}\}\in X_{1}$
and $\{n_{3},n_{4}\}\in X_{2}$) cannot be coded together because
they are forwarded through $n_{5}$. The center node must make a minimum
of two transmissions for every four packets it receives from different
edge nodes in order to ensure that each destination node obtains enough
degrees of freedom to decode. In Fig. \ref{fig:X-OriginalMAC}, when
$P\in[0,\nicefrac{5}{9})$, }NC\textcolor{black}{{} is seen to provide
no additional gains over the use of routing alone since $n_{5}$ can
forward each packet received without the MAC limiting its channel
use. For $P\in[\nicefrac{5}{9},\nicefrac{5}{7}),$ }NC\textcolor{black}{{}
is instrumental in achieving the throughput shown.}

\textcolor{black}{The MAC does not limit channel resources until the
maximum throughput of $S=\nicefrac{5}{7}$ is reached. Assuming symmetric
source loads, this maximum occurs when $\rho_{i}=\nicefrac{1}{7}$
for $i\in[1,5]$ and $p_{T}=\nicefrac{1}{2}\sum_{i=1}^{4}\rho_{i}+\sum_{j=1}^{5}\rho_{j}=1$.
At this maximum, the MAC ensures fairness among all competing nodes
and the throughput saturates for $P_{T}>1$. The non-monotonic behavior
is again due to the fairness aspect of the IEEE 802.11 MAC, and it
is evident that the IEEE 802.11 MAC protocol restricts the total throughput
when the network is saturated.}

\subsubsection{\textcolor{black}{Multi-Packet Reception of Order 2 and 4 (No Network
Coding and $m=2,4$)}}

\textcolor{black}{MPR is similar to the routing case described earlier
except we now allow a maximum of $m$ edge nodes to transmit within
a given time-slot. For $m=2$, the total time used by all of the edge
nodes to transmit their packets to $n_{5}$ is $\nicefrac{1}{2}$
that needed by routing while the center node cannot transmit multiple
packets simultaneously and must transmit each received packet individually.
Using CSMA, which restricts nodes opposite each other to transmit
at the same time, the point at which the protocol saturates for symmetric
source loads occurs when $\rho_{i}=\nicefrac{1}{7}$ for $i\in[1,5]$
and $p_{T}=\sum_{i=1}^{4}\rho_{i}+\nicefrac{1}{2}\sum_{i=1}^{4}+\rho_{5}$.
This maximum, which yields a throughput of $S=\nicefrac{5}{7}$, occurs
when each source has equal loads and is reflected in Fig. \ref{fig:X-OriginalMAC}
as a star. The throughput saturates to the same throughput as routing
for values of $P_{T}>1$ and the gain for $m=2$ is one due to the
suboptimal saturation behavior of the protocol.}

\textcolor{black}{The behavior for $m=4$ is the same as that for
$m=2$ except the maximum of $S=\nicefrac{5}{6}$ occurs when $\rho_{i}=\nicefrac{1}{6}$
and $p_{T}=\sum_{i=1}^{4}\rho_{i}+\nicefrac{1}{4}\sum_{i=1}^{4}\rho_{i}+\rho_{5}=1$.
We allow all of the edge nodes to transmit their packets to $n_{5}$
simultaneously using MPR-adapted CSMA described in Section \ref{sec:Network-models}.
This requires a total of $\nicefrac{1}{6}$ of the time slots. Node
$n_{5}$ then sends each node's packet individually, including its
own, to the intended recipient requiring the remainder of the time-slots
to finish each unicast/broadcast transmission. As $P$ increases,
the MAC limits each node's number of available time-slots and $S$
saturates to $\nicefrac{1}{5}$. Again, the gain in the saturated
region for $m=4$ is equal to the cases of $m=2$ and routing.}

\textcolor{black}{The gain as a result of the use of MPR depends on
an adequate number of source nodes with information to send. If $m$
is greater than the total number of nodes with information to send
(i.e., $m>N$) the MPR gain will be less than when $m\leq N$. In
addition, the achievable gain for implementations using stochastic
message arrival and transmission times will be upper-bounded by the
results shown in this section and lower-bounded by the throughput
for the non-MPR (routing) case seen in Fig. \ref{fig:X-OriginalMAC}.}

\subsubsection{\textcolor{black}{Network Coding with Multi-Packet Reception of Order
2 and 4 ($m=2,4)$}}

\textcolor{black}{The case when MPR is combined with }NC\textcolor{black}{{}
results in further improvement as seen in Fig. \ref{fig:X-OriginalMAC}.
Unlike the case where we considered MPR alone, the order in which
each node transmits and symmetric traffic across the }component\textcolor{black}{{}
are crucial to achieving the maximum throughput. As a result, we continue
to use CSMA to ensure nodes in opposite sets transmit at the same
time so that we both facilitate opportunistic listening and enable
coding opportunities by $n_{5}$.}

\textcolor{black}{For $m=2$, the throughput increases linearly until
it reaches its maximum at $S=1$ when $\rho_{i}=\nicefrac{1}{5}$
for $i\in[1,5]$ and $p_{T}=\nicefrac{1}{2}\sum_{i=1}^{4}\rho_{i}+\nicefrac{1}{2}\sum_{i=1}^{4}\rho_{i}+\rho_{5}=1$.
It then saturates to the }NC\textcolor{black}{{} throughput for $P_{T}>1$.
The decrease in throughput within the saturated region is a direct
result of the IEEE 802.11 MAC which enforces fairness among nodes
and is not a result of the averaging over the stochastic load distributions.
The averaged simulation results and maximum throughput shown in Fig.
\ref{fig:X-OriginalMAC} for $m=2$ is achieved for both unicast and
broadcast traffic when using CSMA to force nodes from different sets
to transmit to $n_{5}$ at the same time. Suppose we use the MPR-adapted
CSMA model so that any two nodes can transmit simultaneously. The
throughput will be the same for unicast traffic, but the broadcast
throughput will be upper bounded by the unicast throughput and lower
bounded by the $m=2$ without }NC\textcolor{black}{{} case. Furthermore,
this shows that the broadcast throughput is dependent on the mechanism
of determining the order of transmissions, such as CSMA, round-robin,
or other similar scheme, within the wireless channel.}

\textcolor{black}{For $m=4$, the maximum unicast throughput of $S=\nicefrac{5}{4}$
is achieved when allowing all four source nodes to transmit to the
center at the same time (i.e., MPR-adapted CSMA is used). The center
node codes a maximum of two native packets together from different
source node sets and transmits two coded packets back to the edge
nodes, including its own uncoded packets, in order to complete all
of the unicast sessions. At the completion of all unicast sessions,
each node still requires a maximum of one additional degree of freedom
to complete the broadcast session. Allowing $n_{5}$ to code all of
the native edge node packets together and send one additional coded
transmission enables each node to extract the required degree of freedom
and obtain the full set of transmitted messages. The maximum throughput
for this case is therefore the same as the case for }NC\textcolor{black}{{}
with $m=2$ and is equal to $S=1$. It is important to note that the
simulated average throughput shown in Fig. \ref{fig:X-OriginalMAC}
for both cases discussed in this sub-section do not reach the maxima
found through analysis, indicated by stars in the figure. This is
due to the stochastic load distribution, which results in asymmetric
traffic among the set of nodes. Should each node have an equal amount
of information to send, the maxima found above will be reached.}

\begin{figure}
\begin{centering}
\textcolor{black}{\includegraphics[width=3.25in]{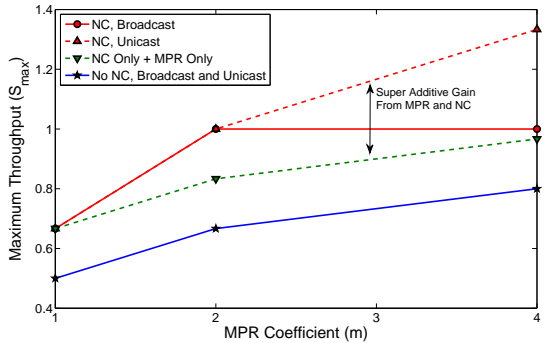}}
\par\end{centering}

\textcolor{black}{\caption{\textcolor{black}{Maximum throughput of a 5-node {}``X'' }component\textcolor{black}{{}
as function of the MPR capability. Super-additive gains are achieved
when using }NC\textcolor{black}{{} in conjunction with MPR. This is
shown by comparing the throughput obtained using both NC and MPR with
the throughput that would be obtained if the gain from using NC alone
is added with the gain obtained from using MPR alone (NC Only + MPR
Only).\label{fig:sat-flow}}}
}
\end{figure}

\textcolor{black}{Fig. \ref{fig:sat-flow} shows a summary of our
analysis by plotting the maximum unicast and broadcast throughput
as a function of the MPR capability. In addition, it illustrates the
super-additive behavior of the throughput when MPR is used in conjunction
with }NC\textcolor{black}{{} by comparing this throughput with the throughput
that would be obtained by adding the individual gains obtained using
MPR and }NC\textcolor{black}{{} separately.}

\subsection{\textcolor{black}{Cross and Partial Topology Component Analysis\label{sub:Cross-Network-Topology}}}

\begin{figure}
\begin{centering}
\textcolor{black}{\includegraphics[width=3.25in]{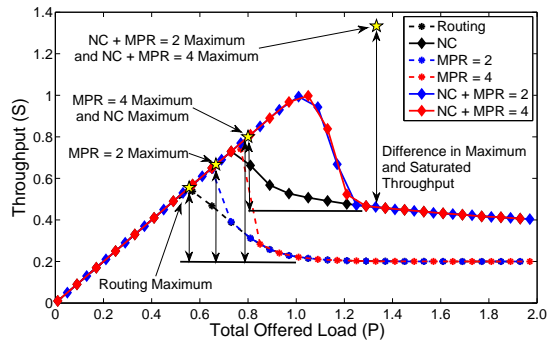}}
\par\end{centering}

\textcolor{black}{\caption{\label{fig:Cross-OriginalMAC}Unicast and broadcast throughput for
a 5-node cross component. The throughput shown is an \emph{average}
over the stochastic packet arrivals and the stars indicate the maximum
throughput which is achieved when there is symmetric full loading
of packets. Each vertical double arrow shows the difference in the
maximum and saturated throughput due to MAC fairness for each case.}
}
\end{figure}

Increasing cooperation between nodes by increasing the number neighbors
each node has results in higher throughput, but there are limitations
to the benefits from increasing the number of overhear/listen edges.
We first consider the cross component shown in Fig. \ref{fig:Network-Topologies}(b),
and conduct a similar analysis performed for the {}``X'' component.
All cases not involving NC are unaffected by the connectivity of the
component, but the maximum throughput for those cases with NC is increased
in the saturated regime. Intuitively, the reason for the increase
in saturated throughput is due to the ability of the center node to
effectively code at most four native packets together. The analysis
is identical to the discussion in Section \ref{sub:X-Topology-Component}
and the results are presented in Fig. \ref{fig:Cross-OriginalMAC}.
The figure show both the analytical results, indicated by the stars,
and the simulation results, indicated by the curves.\textcolor{black}{}

Considering the other possible 5-node components, shown in Fig. \ref{fig:Network-Topologies}(c)
and (d), we find that the addition or removal of an overhear/listen
edge has little impact on the maximum throughput. In the case of the
partial cross component, Fig. \ref{fig:Network-Topologies}(d), the
removal of a single edge results in the maximum throughput found using
the unmodified {}``X'' component. In the case of the partial {}``X''
component, Fig. \ref{fig:Network-Topologies}(c), the gain resulting
from the use of NC is reduced; and as a result, the throughput decreases.
It can be verified using the methods described above that the\textcolor{black}{{}
maximum throughput for the case where }NC\textcolor{black}{{} and $m=2$
is $S=1$ for unicast traffic and $S=\nicefrac{5}{6}$ for broadcast
traffic. This is only a slight reduction in throughput from the unmodified
{}``X'' }component\textcolor{black}{'s throughput. On the other
hand when $m=4$, the maximum is the same as that found for the partial
cross and {}``X'' }component\textcolor{black}{s. Both the partial
cross and partial {}``X'' }component\textcolor{black}{s highlight
that the use of MPR can potentially inhibit the effectiveness of }NC\textcolor{black}{.
Because each node is half-duplex, increasing $m$ restricts each node's
ability to overhear other node's transmissions.}

\section{\textcolor{black}{Improving the MAC Fairness Protocol\label{sec:Improving-the-MAC}}}

\textcolor{black}{The IEEE 802.11 MAC was initially designed for use
in infrastructure wireless networks, yet it is consistently used as
the primary medium access method in ad-hoc, multi-hop networks. Section
\ref{sec:Multi-Packet-Reception-and} showed that the IEEE 802.11
MAC's use in these ad-hoc, multi-hop networks results in the non-monotonic
saturation behavior observed in the COPE experiments \citep{Katti01}.
In this section, we propose an improved MAC approach developed for
use in ad-hoc, multi-hop networks that }\textcolor{black}{\emph{eliminates}}\textcolor{black}{{}
this non-monotonic behavior. Furthermore, our MAC provides fairness
to }\textcolor{black}{\emph{flows}}\textcolor{black}{{} rather than
to }\textcolor{black}{\emph{nodes}}\textcolor{black}{. Our improved
protocol approach allocates resources proportional to the number of
different flows passing through a given node when the network saturates.
While allocating more resources to flows originating at the center
and less resources to flows originated at edge nodes would yield even
higher throughput, our policy ensures that each flow of information
is given the same priority.}

\begin{figure}
\centering{}\textcolor{black}{\includegraphics[width=3.25in]{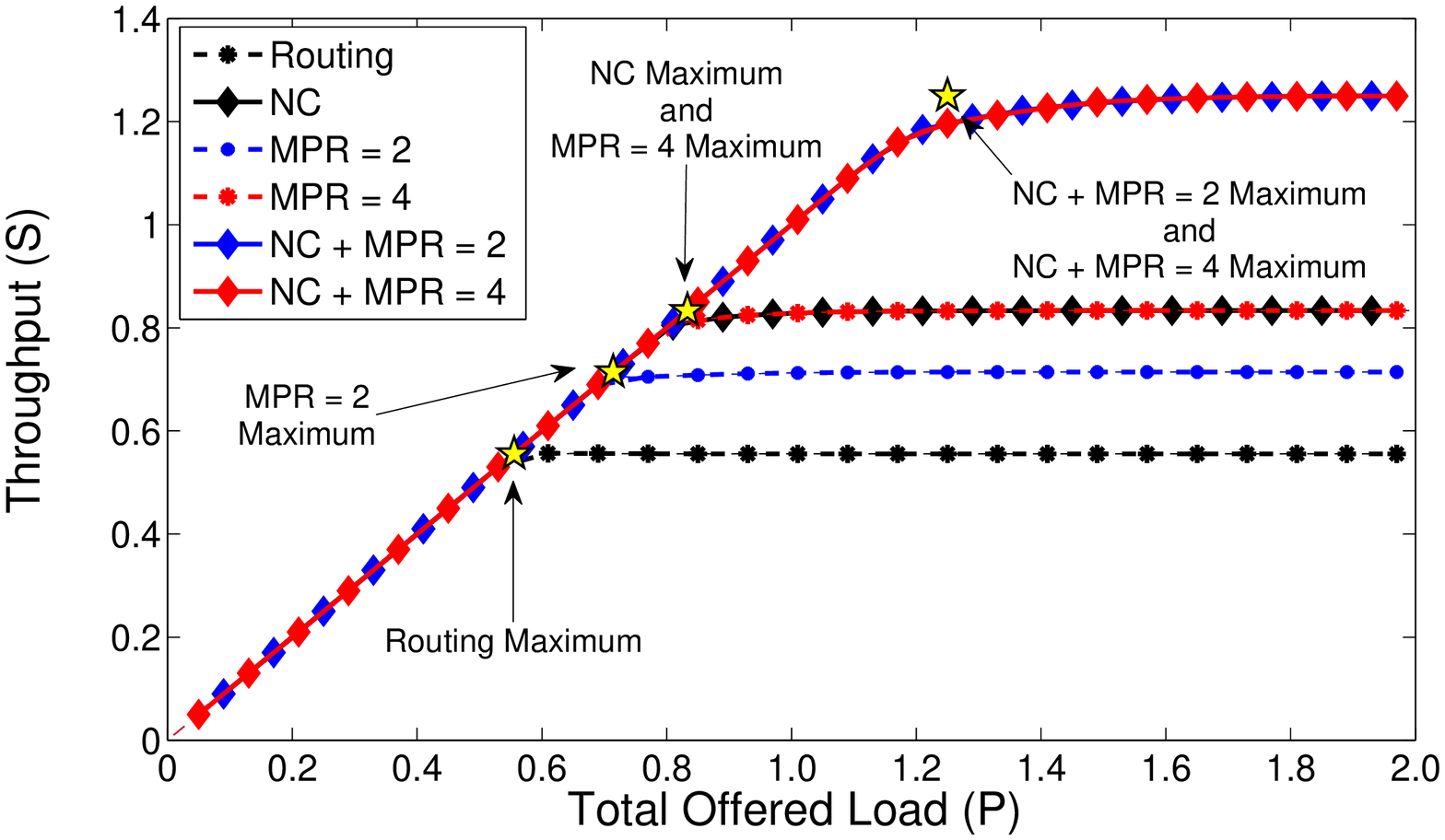}\caption{\label{fig:5Node-Cross_ImprovedMAC}5-Node cross component unicast
and broadcast throughput using the improved fairness protocol. The
throughput shown is an average over the stochastic packet arrivals
and the stars indicate the maximum throughput which is achieved when
there is symmetric full loading of packets.}
}
\end{figure}

\begin{figure}
\begin{centering}
\textcolor{black}{\includegraphics[width=3.25in]{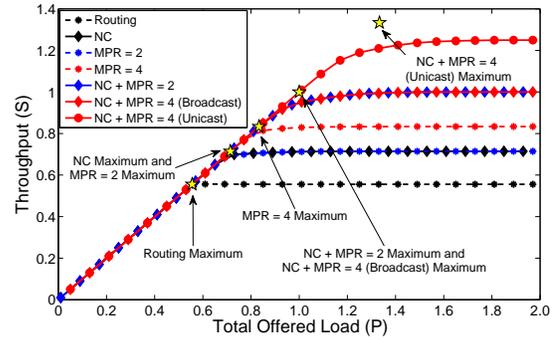}}
\par\end{centering}

\textcolor{black}{\caption{\label{fig:5-Node-X-Unicast}5-Node {}``X'' component throughput
using the improved fairness protocol. The throughput shown is an average
over the stochastic packet arrivals and the stars indicate the maximum
throughput which is achieved when there is symmetric full loading
of packets.}
}
\end{figure}

\textcolor{black}{The allocated number of time-slots each node receives
so that the throughput $S$ is maximized, subject to the flow constraints
and $\sum_{j=1}^{N-1}\nicefrac{s_{j}}{m}+s_{center}=1$, is divided
into the cases below where $s_{j}$ is the fraction of time slots
allocated to each edge node and $s_{center}$ is the fraction of time
slots allocated to the center node. Similar to Section \ref{sec:Multi-Packet-Reception-and},
the throughput $S=s_{center}$ when }NC\textcolor{black}{{} is not used
and $P_{T}\leq1$. When }NC\textcolor{black}{{} is used, the throughput
$S$ is a function of the number of packets that can be effectively
coded together, which is dependent on the MPR coefficient $m$, the
use of CSMA, and the traffic type (unicast or broadcast). The cases
addressed include:}
\begin{itemize}
\item \textsl{\textcolor{black}{Cross Topology Component with Unicast Traffic
or Broadcast Traffic:}}\textcolor{black}{{} Assuming that there are
no constraints on the order in which each node transmits to the center
node, the allocation of resources is the same for both unicast and
broadcast sessions. Without }NC\textcolor{black}{, the center node
will require a number of time slots equal to the number of transmitting
source nodes $N$. With }NC\textcolor{black}{, throughput is maximized
by ensuring the center node codes the maximum number of native packets
together. Generalizing for $N$ and $m$, as well as considering only
integer numbers of time-slots:
\begin{equation}
s_{j}=\begin{cases}
\frac{1}{\lceil(N-1)/m\rceil+N} & \textrm{without NC}\\
\frac{1}{\lceil(N-1)/m\rceil+m_{c}+1} & \textrm{with NC}
\end{cases}\label{eq:cross_bw1}
\end{equation}
and
\begin{equation}
s_{center}\leq\begin{cases}
\frac{N}{\lceil(N-1)/m\rceil+N} & \textrm{without NC}\\
\frac{m_{c}+1}{\lceil(N-1)/m\rceil+m_{c}+1} & \textrm{with NC}
\end{cases}\label{eq:cross_bw2}
\end{equation}
When MPR-adapted CSMA is used, we define $m_{c}=m$ for $m=\{1,2\}$.
When CSMA is used, we define $m_{c}=m-1$ for $m=2$. In addition,
the term $m_{c}=m-1$ for all situations where $m=4$. Furthermore,
eq. (\ref{eq:cross_bw2}) is met with equality if CSMA is used for
$m=1$ and 2 as well as for all cases when $m=4$. Eq. (\ref{eq:cross_bw2})
may be met with inequality when MPR-adapted CSMA is used for $m=2$
since there is a non-zero probability that any given node may miss
a packet from a node in which it can overhear while it is transmitting.
Using a scheme such as CSMA results in a significant throughput gain
for small $N$ but becomes insignificant as $N$ grows.}
\item \textsl{\textcolor{black}{{}``X'' Topology Component: }}\textcolor{black}{The
fraction of time slots $s^{U}$ allocated to each node for unicast
traffic and either the CSMA or MPR-adapted CSMA models is:
\begin{equation}
s_{j}=s_{j}^{U}=\begin{cases}
\frac{1}{\lceil(N-1)/m\rceil+N} & \textrm{without NC}\\
\frac{1}{\lceil(N-1)/m\rceil+\max\left(\mid X_{1}\mid,\mid X_{2}\mid\right)+1} & \textrm{with NC}
\end{cases}\label{eq:x_bw1}
\end{equation}
and 
\begin{equation}
s_{R}=s_{R}^{U}=\begin{cases}
\frac{N}{\lceil(N-1)/m\rceil+N} & \textrm{without NC}\\
\frac{\max\left(\mid X_{1}\mid,\mid X_{2}\mid\right)+1}{\lceil(N-1)/m\rceil+\max\left(\mid X_{1}\mid,\mid X_{2}\mid\right)+1} & \textrm{with NC}
\end{cases}\label{eq:x_bw2}
\end{equation}
When considering broadcast traffic, additional degrees of freedom
must be sent by the center to complete the session. Without }NC\textcolor{black}{,
equations \ref{eq:x_bw1} and \ref{eq:x_bw2} hold. With }NC\textcolor{black}{,
there is a possibility that each destination node will require a maximum
of one additional degree of freedom per node for $m=2$ or three degrees
of freedom per node for $m=4$ when either $\mid X_{1}\mid\geq m$
or $\mid X_{2}\mid\geq m$ and the order of node transmission is not
enforced (i.e., MPR-adapted CSMA). In order to provide these degrees
of freedom, the center node must send send additional coded packets.
The fraction of time-slots each node receives for broadcast traffic,
$s^{B}$, with }NC\textcolor{black}{{} is then bounded by: 
\begin{equation}
{\textstyle s_{j}^{U}\geq s_{j}^{B}\geq\frac{1}{\lceil(N-1)/m\rceil+\max\left(\mid X_{1}\mid,\mid X_{2}\mid\right)+m}}\label{eq:x-broadcast1}
\end{equation}
and 
\begin{equation}
s_{R}^{U}{\textstyle \leq s_{R}^{B}\leq\frac{\max\left(\mid X_{1}\mid,\mid X_{2}\mid\right)+m}{\lceil(N-1)/m\rceil+\max\left(\mid X_{1}\mid,\mid X_{2}\mid\right)+m}}.
\end{equation}
where $s_{j}^{B}$ is maximized when $\left|X_{1}\right|=\left|X_{2}\right|$
and traffic across the center is symmetric. It is minimized when $\left|X_{1}\right|$
and $\left|X_{2}\right|$ differ most and traffic across the center
is asymmetric.}
\end{itemize}
\textcolor{black}{We applied our revised fairness protocol to both
the 5-node cross and {}``X'' }component\textcolor{black}{s using
the same model described in Section \ref{sec:Network-models}. The
throughput, shown in Fig. \ref{fig:5Node-Cross_ImprovedMAC} and \ref{fig:5-Node-X-Unicast},
saturates at the }\textit{\textcolor{black}{maxima}}\textcolor{black}{{}
found in Section \ref{sec:Multi-Packet-Reception-and} for each }component\textcolor{black}{.
As the network saturates, the improved fairness protocol limits each
node's access to the channel. When each node's load is greater than
the limit imposed by the protocol, the total throughput will saturate
at the maxima. As the simulation results represented by the curves
in both figures indicate, the maxima may not be reached due to asymmetry
in each node's load and is the reason why the average throughput shown
in Fig. \ref{fig:5Node-Cross_ImprovedMAC} and \ref{fig:5-Node-X-Unicast}
do not initially saturate at their maxima. As the network initially
saturates, some nodes will have higher loads than others resulting
in a lower throughput than when each node has the same load which
occurs as $P$ increases towards infinity.}
\textcolor{black}{}

\section{\textcolor{black}{Performance of MPR and Network Coding with Asymmetric
Traffic\label{sec:Asymmetric-Traffic}}}

\textcolor{black}{The performance of }NC\textcolor{black}{{} and MPR
in networks with bottlenecks is highly dependent on the symmetry of
traffic across the bottleneck. Situations in which the traffic is
approximately symmetric, or equal, across the bottleneck maximizes
the performance gains provided by both }NC\textcolor{black}{{} and MPR
as shown by the stars in Fig. \ref{fig:X-OriginalMAC}, \ref{fig:Cross-OriginalMAC},
\ref{fig:5Node-Cross_ImprovedMAC}, and \ref{fig:5-Node-X-Unicast}.
The curves in each figure show that the maxima are not initially reached
since each curve represents an average over the instantaneous asymmetries
in traffic. For the purposes of analyzing the effects of asymmetric
traffic, the {}``X'' }component\textcolor{black}{{} is used as the
primary }component\textcolor{black}{{} in our analysis since its limitations
from the reduced number of nodes any given edge node can overhear
compounds the effects of asymmetric traffic on network throughput.
In addition, we define the asymmetry ratio $\nu$ as:
\begin{equation}
\nu=\frac{\sum_{i\in X_{2}}k_{i}}{\sum_{j\in X_{1}}k_{j}},
\end{equation}
 where $k_{i}$ and $k_{j}$ are the number of packets that each node
$i\in X_{2}$ and $j\in X_{1}$, respectively, needs to send to a
given node on the opposite side of the relay.}

\textcolor{black}{Two different asymmetry scenarios are addressed.
The first addresses the effects of asymmetry with a MAC that limits
the transmissions of nodes from the same set (i.e., nodes within the
same set do not transmit at the same time unless the degree of MPR
requires that they do so). In this scenario, both the effectiveness
of NC and MPR is diminished as $\nu$ increases. When $m=2$, only
a single node from a set will transmit in a time-slot, corresponding
to CSMA. As traffic becomes more asymmetric, one set of nodes will
eventually run out of data and the other set will be forced to continue
sending data to the relay one node at a time. For $m=4$, two nodes
from the same set will transmit in the same time-slot since the }component\textcolor{black}{{}
contains only two sets of nodes, which corresponds to MPR-adapted
CSMA. When NC is used, the limitations induced by the component force
the center node to transmit packets unencoded when $\nu$ is large.
For example, as $\nu$ increases, the center node will run out of
data from different sets to code together. As a result, each packet
that needs to be relayed must be forwarded individually to ensure
that the necessary degrees of freedom are exchanged.}
\begin{figure}
\begin{centering}
\textcolor{black}{\includegraphics[width=3.25in]{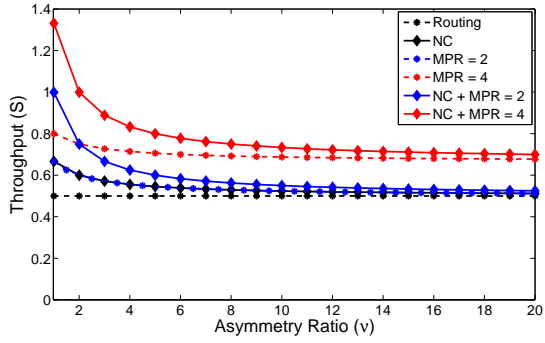}}
\par\end{centering}

\textcolor{black}{\caption{Throughput of an {}``X'' component as a function of the asymmetry
ratio with an offered load of 1 when CSMA is used to limit transmission
order.\label{fig:Asymmetry}}
}
\end{figure}

\textcolor{black}{Fig. \ref{fig:Asymmetry} shows how asymmetric traffic
from different sets affects network throughput. For the case when
$m=2$ with or without }NC\textcolor{black}{, the throughput is maximized
when traffic is perfectly symmetric but saturates to the throughput
obtained for the routing case. Intuitively, this can be seen by limiting
all traffic so that it originates from a single set of nodes. The
MAC will restrict transmission from each node to the center, which
eliminates the gain resulting from the use of MPR; and the center
node must send each relayed packet uncoded to the next hop, which
eliminates the gain resulting from }NC\textcolor{black}{. The case
for $m=4$ is similar to that of the $m=2$ case except that only
a maximum of two nodes in a set are allowed to transmit in the same
time slot. The throughput will saturate for large $\nu$ to the maximum
$m=2$ throughput of $\nicefrac{2}{3}$.}

The second scenario involves the use of a MAC that does not limit
the number of nodes that simultaneously transmit in either set $X_{1}$
or $X_{2}$.\textcolor{black}{{} The MAC allows nodes within the same
set to take advantage of MPR and does not restrict multiple nodes
from sending to the relay in a given time slot. If only nodes within
the same set have data to send, the MAC allows for up to $m$ nodes
to send their respective packets to the relay. In this scenario, the
effectiveness of MPR is not diminished since MPR can be fully utilized
regardless of where the traffic is originates. This results in a constant
throughput, independent of $\nu$, of $S=\nicefrac{2}{3}$ for the
$m=2$ case and $S=\nicefrac{4}{5}$ for the $m=4$ case. On the other
hand, the effectiveness of }NC\textcolor{black}{{} still decreases as
$\nu$ increases. Similar to the first scenario, the throughput for
each case involving NC will saturate to the routing or MPR only throughput
for each case involving NC as $\nu$ increases.}

\textcolor{black}{This section emphasizes that implementing a MAC
that allows for the full employment of MPR provides significant throughput
gains over a more restrictive MAC, such as one that uses a CSMA scheme.
Finally, it is important to note, that in the presence of erasures
the potential gains are significant even with asymmetric traffic.
While }NC\textcolor{black}{{} may not necessarily increase throughput
under highly asymmetric data, the }NC\textcolor{black}{{} gain will
manifest itself when recovering from packet erasures.}

\section{\textcolor{black}{Performance of Network Coding and MPR with Large
N\label{sec:Performance-of-large-N}}}

\textcolor{black}{The gain provided by the use of MPR and }NC\textcolor{black}{{}
is dependent on the number of transmitting nodes $N$ within the }component\textcolor{black}{.
While the gain manifests itself in the throughput of each canonical
}component\textcolor{black}{, the major benefit is realized in the
delay, or time it takes to complete all flows. For purposes of illustration,
we restrict our analysis to the cases in which we have a restrictive
MAC which uses CSMA, symmetric traffic across each }component\textcolor{black}{,
and the improved fairness protocol. Using eq. (\ref{eq:cross_bw1})
- (\ref{eq:x_bw2}), relaxing the integer constraints, and assuming
an equal number of nodes in each set within the {}``X'' }component\textcolor{black}{,
we take the limit of the throughput for each canonical }component\textcolor{black}{:
\begin{equation}
\lim_{N\rightarrow\infty}S_{Cross}=\begin{cases}
\frac{m}{m+1} & \textrm{without NC}\\
m & \textrm{with NC}
\end{cases}
\end{equation}
\begin{equation}
\lim_{N\rightarrow\infty}S_{X}=\begin{cases}
\frac{m}{m+1} & \textrm{without NC}\\
\frac{2m}{m+2} & \textrm{with NC}
\end{cases}
\end{equation}
It is clear from the above results that the gain has a dependency
on the connectivity of the network. As the network becomes more connected,
the interaction between }NC\textcolor{black}{{} and MPR combine to create
gains that are super-additive.}

\begin{figure}
\begin{centering}
\textcolor{black}{\includegraphics[width=3.25in]{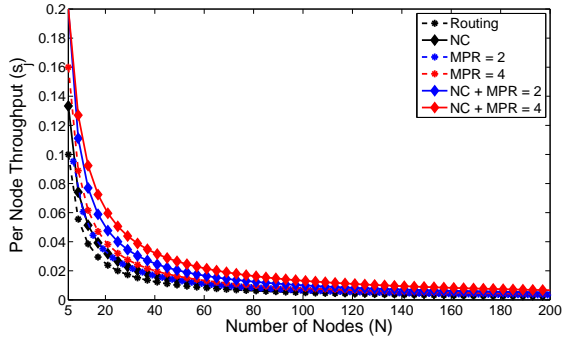}}
\par\end{centering}

\textcolor{black}{\caption{Throughput per node of the {}``X'' component for large N using the
improved fairness protocol.\label{fig:Throughput-per-node}}
}
\end{figure}

\textcolor{black}{Considering the per-node throughput $S_{Node}=s_{j}$
for $j\in[1,N]$, we see from eq. (\ref{eq:x_bw1}) that the throughput
for both the original IEEE 802.11 MAC and improved MAC scales on the
order of $\nicefrac{1}{N}$. Fig. \ref{fig:Throughput-per-node} shows
the $\nicefrac{1}{N}$ per-node throughput behavior for the {}``X''
}component\textcolor{black}{, using the improved MAC, as a function
of the number of nodes. As expected, the per-node throughput asymptotically
approaches zero as $N$ grows. While there are gains from MPR and
}NC\textcolor{black}{{} for moderately sized networks (i.e., $N=[5,100]$)
the throughput gains are limited for larger ones.}

\begin{figure}
\begin{centering}
\textcolor{black}{\includegraphics[width=3.25in]{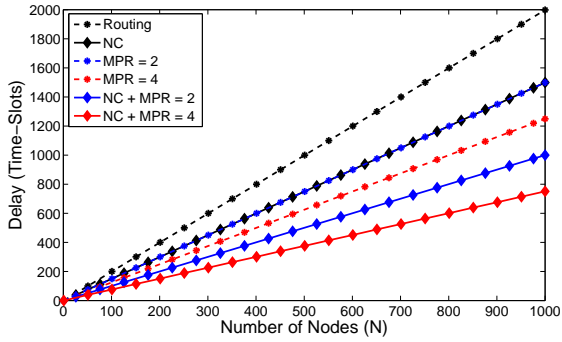}}
\par\end{centering}

\textcolor{black}{\caption{Time to complete all flows if each source has only a single packet
to send using the improved fairness protocol.\label{fig:Delay}}
}
\end{figure}

\textcolor{black}{On the other hand, there are significant gains from
MPR and }NC\textcolor{black}{, while using the improved MAC, when
considering the }\textit{\textcolor{black}{delay}}\textcolor{black}{,
or total time to complete all sessions. We evaluate the delay by distributing
a single packet to each node and determine the time it takes for all
packets to reach their intended destinations. Fig. \ref{fig:Delay}
shows the total time to complete all flows within an {}``X'' }component\textcolor{black}{{}
as $N$ grows. It can be verified from Fig. \ref{fig:Delay} that
the delay gains for the MPR with $m=2$ or $m=4$ and }NC\textcolor{black}{{}
cases are approximately $2$ and $\nicefrac{8}{3}$ respectively for
large $N$.}
\textcolor{black}{}

\section{MPR Limited to Central Node\label{sec:Limited-MPR}}

Since implementing MPR in a system may be a difficult and costly upgrade,
we now look at the throughput gains if we target strategic nodes for
implementing MPR and leave the rest without the capability. The limitation
of not having MPR at each edge node, as would be expected, reduces
the effectiveness of opportunistic NC and limits the total number
of packets that the center node can code together.

We continue to use CSMA as explained in Section \ref{sec:Network-models}.
We deterministically distribute an equal number of packets to each
node and calculate the throughput using as the number of transmitting
nodes $N$ increases towards infinity. We further assume that each
node has the ability to capture a packet. That is, if multiple transmissions
occur in a given time-slot, a node without MPR will receive one transmission
without error and treat the remaining transmissions as noise. If capture
is not feasible, the NC with MPR gain will equal the NC alone gain
for components such as the cross. The NC with MPR gain for less connected
topologies, in contrast, will be higher depending on the implementation
of the MAC since the topology limitations decrease the probability
of two nodes' transmissions conflicting. For ease of further explanation,
we will assume in the remainder of the paper that every node has the
ability to capture packets.

The number of additional packets that the center node must send when
each edge node does not have MPR is dependent on $m$. Limiting MPR
to the center node essentially splits a component into $m$ disjoint
sets where all edge nodes in a set are fully connected and each node
is connected to the center. An MPR of $m=2$ will result in two disjoint
sets that requires the center node to send $\left\lceil \nicefrac{(N-1)}{2}\right\rceil +1$
degrees of freedom to each edge node in order to complete all unicast
and broadcast sessions. The first term in this equation is the number
of transmissions needed to relay all traffic from each of the edge
nodes and the second is the number of transmissions needed to send
the center's own traffic. In the case of the {}``X'' component,
the division has already been performed as a result of the topology
configuration so the throughput is the same as that found in Section
\ref{sec:Improving-the-MAC}. The throughput for the cross component
becomes the same as that of the {}``X'' component as a result of
the limited implementation of MPR. An MPR of $m=4$ results in four
disjoint sets that requires the center node to send $\left\lceil \nicefrac{(N-1)}{2}\right\rceil +1$
degrees of freedom to the set of edge nodes to complete all unicast
sessions and $\left\lceil \nicefrac{3(N-1)}{4}\right\rceil +1$ degrees
of freedom to each edge node to complete the broadcast session. Within
both components, the result of increasing $m$ is offset by the requirement
of the center node to send additional degrees of freedom. The broadcast
throughput for both components becomes upper bounded by the throughput
of the {}``X'' component when using both NC and $m=2$; and the
unicast throughput for NC with $m=4$ is upper bounded by $\nicefrac{4}{3}$
for both components.

\begin{figure}
\centering{}\includegraphics[width=3.5in]{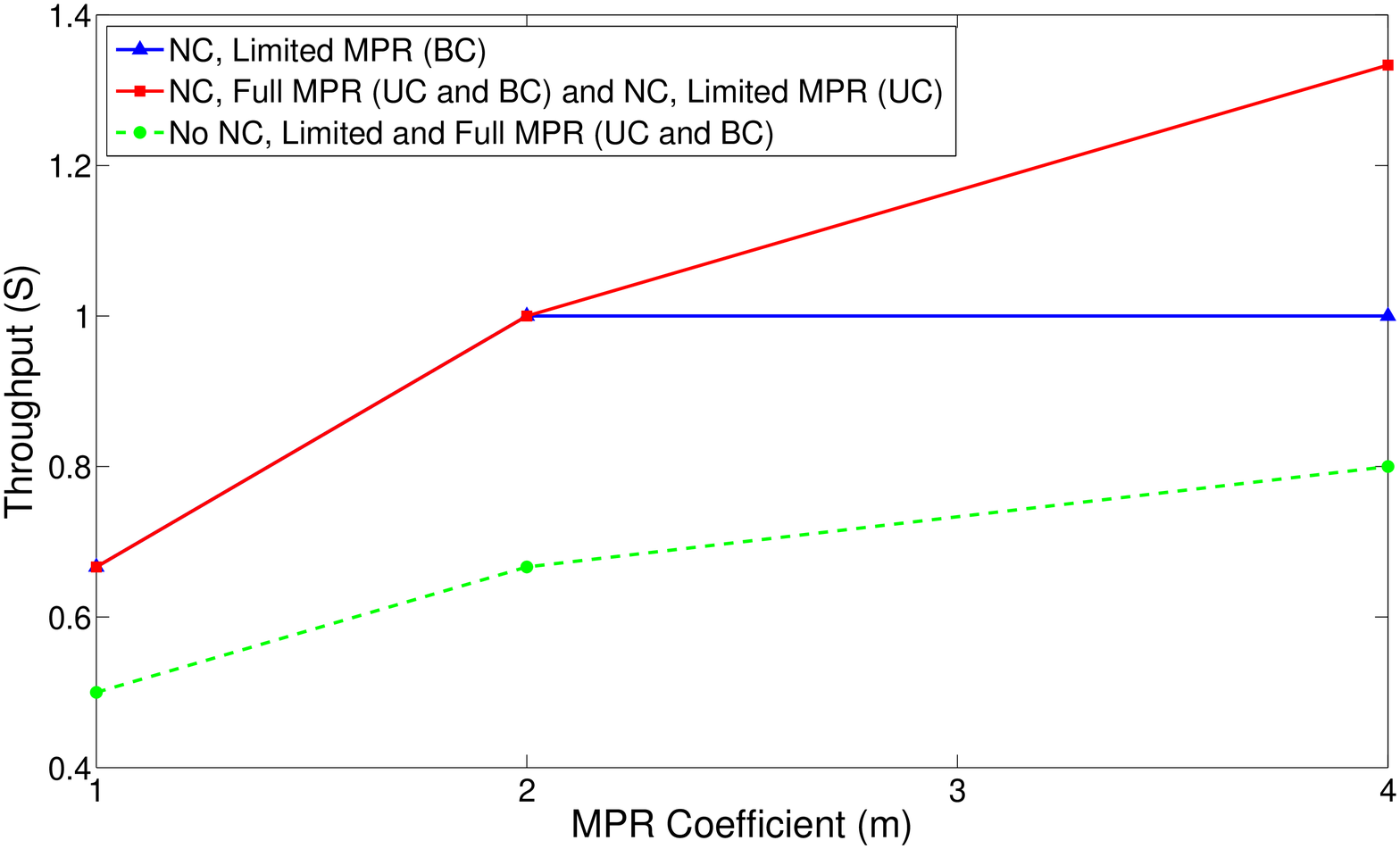}\caption{Unicast (UC) and broadcast(BC) throughput as $N\rightarrow\infty$
for the {}``X'' component with MPR implemented at each node (full
MPR) and MPR implemented at only the center node (limited MPR) under
fully saturated and symmetric loading.\label{fig:Limited-MPR}}
\end{figure}

The cases where NC is not used are unaffected by limiting MPR to the
center node only. Since the center must forward all packets individually,
it inherently communicates all of the necessary degrees of freedom
to each of the edge nodes. These results are displayed in Fig. \ref{fig:Limited-MPR}
for the {}``X'' component as we let $N$ increase towards infinity
and consider fully saturated, symmetric loading. This figure was generated
using a similar analysis as performed in Section \ref{sec:Performance-of-large-N}.
When considering the cross component, the throughput with limited
MPR is the same as that shown in the above figure, but the NC + MPR
throughput with MPR implemented at each node is $S=1$, $S=2$, and
$S=4$ for $m=1$, $m=2$, and $m=4$ respectively. This shows that
there are significant throughput gains when considering NC with MPR
for topologies similar to the {}``X'' component, but little for
topologies that are more connected or for larger values of $m$.

\section{\textcolor{black}{Conclusion\label{sec:Conclusion}}}

\textcolor{black}{We have provided a lower bound to the gains in total
throughput from MPR and }NC\textcolor{black}{{} for }component\textcolor{black}{s
that create traffic bottlenecks in large networks. We provided an
analysis of the total throughput and showed that the effectiveness
of }NC\textcolor{black}{{} is highly dependent on the use of MPR, and
that the combined use of MPR and }NC\textcolor{black}{{} results in
}\textcolor{black}{\emph{super-additive}}\textcolor{black}{{} gains.
In addition, we evaluated the fairness imposed by the IEEE 802.11
MAC and showed that the NC + MPR gain at saturation is not maximized.
We argued that while the current IEEE 802.11 MAC is fair to }\textit{\textcolor{black}{nodes}}\textcolor{black}{,
it is inherently unfair to }\textit{\textcolor{black}{flows}}\textcolor{black}{{}
of information in multi-hop networks. We further generalized each
scenario for both unicast and broadcast traffic.}

\textcolor{black}{We then used our simple, validated model to design
a new MAC approach, in conjunction with MPR and NC, that cooperatively
allocates channel resources by providing a greater proportion of resources
to bottle-necked nodes and less to source nodes. The new MAC ensures
fairness among information }\textit{\textcolor{black}{flows}}\textcolor{black}{{}
rather than }\textit{\textcolor{black}{nodes}}\textcolor{black}{{} through
the cooperative allocation of bandwidth between the set of edge nodes
and the center node. Furthermore, the new MAC ensures that each node
is able to access the channel in contrast to the IEEE 802.11 MAC which
has a tendency to starve some nodes under high loads. Our proposed
approach, specifically designed for networks using }NC\textcolor{black}{{}
and MPR, shows a significant increase in the achievable throughput
of as much as 6.3 times the throughput when neither }NC\textcolor{black}{{}
nor MPR is used in similar networks. While only four specific 5-node
canonical topology components and their extensions to $N$ nodes were
addressed, these components serve as a basis for further investigation
on how channel resource allocation should be performed in larger,
more complex networks.}

\textcolor{black}{In addition, we analyzed the scalability of the
canonical topology components. We showed that the gains provided by
the use of MPR and }NC\textcolor{black}{{} are highly dependent on the
connectivity of the network. While the asymptotic per-node throughput
is not large, the asymptotic gains in the delay are substantial. We
further showed that asymmetric loads across a bottleneck can impact
network performance when using both }NC\textcolor{black}{{} and MPR,
although }NC\textcolor{black}{{} and MPR still provide significant gains
for low to medium asymmetric loads. Finally, we showed that limiting
the distribution of the MPR capability to only a subset of nodes within
a network can result in a drastic reduction in performance for some
canonical topologies. Less connected components such as the {}``X'',
which are much more likely to be physically realizable in contrast
to the cross component, are less affected by limiting the implementation
of MPR to only a subset of nodes. In contrast, components that are
more connected, such as the cross, lose much of the throughput gain
resulting from combining NC with MPR. All of the analyses outlined
in this paper show that the cooperative use of MPR, }NC\textcolor{black}{,
and MAC in a given network is critical to achieving the maximum gain.\bibliographystyle{IEEEtran}
\bibliography{JSACBib}
}
\end{document}